\documentclass[sigconf]{acmart}

\AtBeginDocument{%
  }

\copyrightyear{2026}
\acmYear{2026}
\setcopyright{cc}
\setcctype{by}
\acmConference[CHI '26]{Proceedings of the 2026 CHI Conference on Human Factors in Computing Systems}{April 13--17, 2026}{Barcelona, Spain}
\acmBooktitle{Proceedings of the 2026 CHI Conference on Human Factors in Computing Systems (CHI '26), April 13--17, 2026, Barcelona, Spain}
\acmDOI{10.1145/3772318.3791154}
\acmISBN{979-8-4007-2278-3/2026/04}

\usepackage{geometry}
\usepackage{makecell}

\begin{document}

\title[Understanding Agency of Home-Based Care Patients]{``I Choose to Live, for Life Itself'': Understanding Agency of Home-Based Care Patients Through Information Practices and Relational Dynamics in Care Networks}

\author{Sung-In Kim}
\authornote{These authors contributed equally to this research.}
\orcid{0000-0001-7627-2756}
\affiliation{%
  \institution{Department of Psychiatry}
  \institution{Seoul National University Bundang Hospital}
  \city{Seongnam}
  \country{Republic of Korea}
}
\email{sunginkim@snu.ac.kr}

\author{Joonyoung Park}
\authornotemark[1]
\orcid{0009-0008-5003-6673}
\affiliation{%
  \institution{Department of Industrial Design}
  \institution{KAIST}
  \city{Daejeon}
  \country{Republic of Korea}
}
\email{joonyoung@kaist.ac.kr}

\author{Bogoan Kim}
\orcid{0000-0002-9083-1128}
\affiliation{%
  \institution{School of Information and Communication Engineering}
  \institution{Chungbuk National University}
  \city{Cheongju}
  \country{Republic of Korea}
}
\email{bogoan@cbnu.ac.kr}

\author{Hwajung Hong}
\authornote{Corresponding author}
\orcid{0000-0001-5268-3331}
\affiliation{%
  \institution{Department of Industrial Design}
  \institution{KAIST}
  \city{Daejeon}
  \country{Republic of Korea}
}
\email{hwajung@kaist.ac.kr}


\newcommand{\revise}[1]{%
  \textcolor{red}{\fcolorbox{red}{red}{\textcolor{white}{R}}~#1}%
}






\begin{abstract}
Home-based care (HBC) delivers medical and care services in patients' living environments, offering unique opportunities for patient-centered care. However, patient agency is often inadequately represented in shared HBC planning processes. Through 23 multi-stakeholder interviews with HBC patients, healthcare professionals, and care workers, alongside 60 hours of ethnographic observations, we examined how patient agency manifests in HBC and why this representation gap occurs. Our findings reveal that patient agency is not a static individual attribute but a relational capacity shaped through maintaining everyday continuity, mutual recognition from care providers, and engagement with material home environments. Furthermore, we identified that structured documentation systems filter out contextual knowledge, informal communication channels fragment patient voices, and doctor-centered hierarchies position patients as passive recipients. Drawing on these insights, we propose design considerations to bridge this representation gap and to integrate patient agency into shared HBC plans.
\end{abstract}

\maketitle

\section{Introduction}

Home-based care (HBC) represents an integrated approach that delivers medical and care services within patients' familiar home environments, specifically targeting individuals who face difficulties accessing hospital or institutional care due to cognitive or physical limitations \cite{fabius_caregiving_2023}. By operating within patients' living spaces rather than clinical settings, HBC naturally accommodates individual characteristics, living environments, and social networks, offering unique opportunities for patient-centered care (PCC) through continuous connections within care networks \cite{hughes_effectiveness_2000}.

Given the nature of these services, HBC recipients are predominantly older adults with multiple chronic diseases (e.g., hypertension, diabetes) or disabilities (e.g., dementia, spinal cord injury) \cite{schmidt-mende_profiling_2024, mondor_multimorbidity_2017, mello2023}. HBC has demonstrated various positive health outcomes, including improved medication adherence, chronic disease prevention, and reduced depression \cite{bayen_chronic_2024, eltaybani_toward_2023, lizano-diez_impact_2022, ergin_effect_2022, ramli_effectiveness_2024}. Building on these advantages, many countries are expanding their HBC services as populations age and the number of patients with multiple chronic diseases and disabilities continues to grow  \cite{jacobs_patient_2019, irani_care_2007, lee_kun-sei_home_2023, llena2025countries}.

HBC extends beyond the delivery of medical services, functioning as a sociotechnical practice that negotiates treatment and care into patients' daily lives \cite{livingston2017dementia}. In this situated context, patients are not passive recipients of clinical interventions but active agents whose capacity to shape care is co-constructed through their domestic environments and relationships with care providers \cite{prah_ruger_health_2010}. Accordingly, rather than viewing patient agency through the traditional lens of individual decision-making capacity, it can be understood through a care ethics perspective as a relational process of navigating interdependence to preserve identity \cite{sherwin1998politics, lonkila2021care}. Therefore, to realize a truly PCC, it is vital to examine not only how this agency is expressed but also how it is mediated or obscured by the care networks \cite{entwistle_treating_2013, lai2022, elwyn2017}.

Prior HCI research has extensively examined the sociotechnical dimensions of healthcare, investigating how coordination artifacts mediate clinical work in hospitals \cite{bardram2005, reddy2001coordinating} and how domestic technologies reconfigure care relationships in the home \cite{soubutts2021aging, branham2015collaborative}. Building on this foundation, scholars have proposed designs to support communication and mutual understanding among care stakeholders \cite{hsu2024, yamashita_how_2018, sun2023data}. However, existing studies have yet to fully address the distinct infrastructural dynamics of HBC, where professional medical protocols intersect with the personal lifeworlds of patients. Specifically, there is limited empirical insight into how the distributed information practices of these multi-stakeholder networks systematically structure and potentially constrain the visibility and legitimacy of patient agency.


With this background, we aimed to understand how patient agency manifests in HBC networks. We conducted 23 multi-stakeholder interviews with HBC patients, healthcare professionals, and care workers, along with 60 hours of ethnographic observations. 
Our findings reveal that patient agency in HBC operates as a mechanism to preserve a sense of self within dependence, relying on mutual recognition from care providers and the material home environment in mediating agency. Simultaneously, we found critical barriers where intimate relationships sometimes created tensions that constrained expression. Furthermore, current sociotechnical infrastructures of HBC reproduced doctor-centered power dynamics, creating a representation gap in which patient inputs were systematically filtered out of shared care planning.

This study contributes theoretically and empirically in three ways. First, we reveal and articulate a relational account of patient agency specific to HBC contexts, demonstrating how HBC patient agency is co-constructed through maintaining everyday continuity, achieving mutual recognition, and engaging with material home environments. Second, we introduce the representation gap as an analytic construct that reveals how information practices and relational dynamics in HBC networks systematically filter patient expressions. Third, we derive design considerations that operationalize these insights to bridge the representation gap in HBC coordination.


\section{Related Work}
In this section, we situate our work within two intersecting domains essential for understanding the dynamics of patient agency in HBC. First, we examine the conceptualization of patient agency within the HBC context, tracing the shift from traditional individualistic definitions to relational frameworks in care ethics and HCI that view agency as co-constructed through social and material environments. Second, we analyze the role of information practices and coordination artifacts in distributed healthcare networks, exploring how documentation and communication infrastructures mediate the visibility and legitimacy of patient perspectives. These bodies of work provide a theoretical foundation to understand patient agency not as an inherent trait, but as a capacity constituted through the complex interplay of relationships, domestic spaces, and sociotechnical systems.

\subsection{Patient Agency within the HBC Contexts}
HBC is an integrated care model in which patients receive medical and care services in their own homes \cite{fabius_caregiving_2023}. Globally, HBC is expanding primarily for elderly individuals with mobility limitations, chronic disease patients, people with disabilities, and end-of-life patients \cite{jacobs_patient_2019, irani_care_2007, lee_kun-sei_home_2023}. Previous research on HBC presents consistent evidence that HBC contributes to improved medication adherence \cite{eltaybani_toward_2023}, chronic disease prevention \cite{bayen_chronic_2024}, nutritional status improvement, and depression reduction \cite{lizano-diez_impact_2022, ergin_effect_2022, ramli_effectiveness_2024, yaron2024constant}. At the same time, HBC can reduce overall healthcare costs and achieve significantly higher patient and family satisfaction compared to institution-based care \cite{caplan2012meta, shepperd2008admission, levine2018}.

The most distinctive characteristic of HBC is that care is provided in patients' homes, which are not merely physical places but lifeworlds where patients' identities, values, and lifestyles are embodied \cite{livingston2017dementia}. This spatial characteristic transforms care and medical practice from episodic encounters into ongoing work situated within patients' daily routines, social relationships, and material environments \cite{arab2022nurses, klein2017overview}. Unlike institutional settings, these sustained, context-rich interactions create conditions where patient agency becomes more visible and consequential, as care networks continuously navigate patients' preferences, priorities, and everyday practices \cite{forsgren2021interactional}. Understanding how patient agency is expressed, recognized, and integrated within HBC environments therefore becomes central to realizing truly patient-centered HBC \cite{entwistle_treating_2013, lai2022, elwyn2017}.

Broadly, agency refers to the capacity to control one's actions and shape the surrounding world \cite{bandura1999social, haggard2017sense}. In healthcare contexts, \textit{patient agency} is typically defined as the ability of patients to make decisions regarding their health, weigh treatment options within the context of their lives, and actively shape their care pathways \cite{prah_ruger_health_2010}. This conceptualization draws primarily from individualistic frameworks that position independent decision-making capacity and self-determination as the cornerstones of agency \cite{ryan2000self}. From this perspective, agency is viewed as an inherent attribute of individuals, exercised through rational choice and autonomous action \cite{acke2022one}.

However, this individualistic framing has faced challenges when applied to care contexts where patients experience physical or cognitive constraints \cite{collopy1988autonomy}. There is concern that positioning independence as a prerequisite for patient agency may inadvertently frame vulnerability and dependence as deficits, which may diminish one's capacity for meaningful participation in care \cite{guldenpfennig2019, collopy1995power}. 

Drawing on a rich tradition of feminist scholarship, care ethics reorients this conventional understanding of patient agency by centering interdependence and relationality in the practice of care \cite{held2005ethics}. Rather than viewing patient agency as individual self-sufficiency, this perspective understands it as emerging through ongoing negotiations and mutual dependence within networks of care relationships \cite{sherwin1998politics, lonkila2021care}. According to Tronto \cite{tronto2020moral}, care is understood not as a unidirectional provision of services but as dynamic work involving multiple dimensions: recognizing needs, taking responsibility, providing care, and responding to the care process. 
This view extends further to recognize that patient agency is co-constructed through care relationships with human and non-human stakeholders, such as care participants, technologies, and their living environments \cite{mol2008logic}.

The relational understanding of patient agency has found productive application within HCI/CSCW research, where scholars examine how technologies and care relationships shape the conditions under which patient agency is expressed and recognized. Research in domestic care settings has revealed that assistive technologies, such as stairlifts, do not simply augment individual capabilities but also reconfigure agency, responsibilities, and emotional relationships across entire household networks \cite{soubutts2021aging}. Similarly, while self-care technologies appear to strengthen patient agency, their design often privileges medical system needs and professional perspectives, potentially constraining patients' capacity to explore problems independently and participate meaningfully in decision-making \cite{nunes2019agency}. 
These studies illustrate that patient agency is inherently collaborative, requiring the negotiation of interdependent needs between care networks, technological artifacts, and the materiality of the domestic setting \cite{branham2015collaborative, mol2008logic}.

Building on this relational framework, HBC environments present a critical yet under-explored context for examining patient agency. While HCI has extensively investigated domestic care, the specific dynamics of HBC, where professional medical protocols intersect with the private lifeworld of the home, create unique sociotechnical tensions that warrant closer scrutiny. Therefore, this study aims to unravel the complexities of patient agency in HBC by exploring not only how it is expressed and overlooked in care relationships, but also how these relationships are structured and recognized by the sociotechnical systems that facilitate care coordination. By examining the interplay between patient expressions and the responsive practices of care networks, this research provides empirical insights into how patient agency is co-constructed in the wild, offering a foundation for designing more agency-affirming HBC technologies.

\subsection{Information Practices and Artifacts Shaping Patient Agency}
Healthcare coordination relies on complex information practices that extend far beyond simple data exchange. Clinical work is fundamentally mediated by material artifacts and documentation systems that shape what information becomes visible, to whom, and in what form. These artifacts not only transmit information but actively construct and constrain how different stakeholders understand and respond to patient needs \cite{bardram2005}.

The infrastructural nature of these information practices becomes particularly evident when examining documentation systems. While designed to ensure safety and efficiency, healthcare documentation systems embody particular institutional logics and values. These systems can encode specific views of legitimate work and distribute power unevenly among stakeholders \cite{pine2014}. This challenge is more pronounced when diverse professionals collaborate on patient care. Although they rely on the same information, they interpret it through distinct professional lenses \cite{reddy2001coordinating}. In this context, information does not function as a uniform representation of reality but as a resource that is continually debated and resolved across divergent work practices \cite{schmidt1992taking}.

Collectively, this body of work illustrates that healthcare decision-making emerges through distributed practices involving artifacts, protocols, organizational structures, and collaborative routines \cite{lee2006human}. This sociotechnical perspective has profound implications for understanding patient agency. When documentation systems privilege certain types of knowledge over others, or when coordination artifacts fail to capture the contextual richness of patient experiences, expressions of agency become invisible or distorted within care networks \cite{willis2019m}. Patient agency, in turn, becomes conditional on whether and how it can be successfully mediated through these infrastructural formats.

Recognizing these challenges, HCI researchers have explored various approaches to support coordination in care networks and strengthen patient participation. Studies have examined how technologies can facilitate reflection on caregiving roles and improve mutual understanding among stakeholders \cite{hsu2024, yamashita_how_2018}, as well as how communication and data-sharing platforms can bridge professional and lay boundaries in care \cite{doyle2019, Curtis2025IUV, sun2023data}. Additionally, research indicates that technologies visualizing communication patterns and care activities can make the often invisible contributions of diverse stakeholders more visible and actionable \cite{bascom_designing_2024, mondor_multimorbidity_2017}. These inquiries demonstrate how sociotechnical design can actively reshape clinical collaboration to create more inclusive spaces for participation.

However, the distinctive context of HBC presents additional complexities that need further investigation. In HBC, care unfolds across distributed networks where each stakeholder operates with different temporal rhythms, uses distinct documentation practices, and holds varying definitions of what information matters for care planning \cite{scheer2024designing}. Furthermore, patients' homes contain rich contextual information that extends beyond conventional clinical data, yet this contextual richness must somehow be captured, interpreted, and integrated across a fragmented care network \cite{joo2023fragmented}. 

Drawing on this line of research, we examine how these dynamics specifically shape the sociotechnical conditions for patient agency in HBC. We analyze how material artifacts, organizational structures, and relational dynamics within HBC networks determine what becomes visible and actionable in the shared care planning process, ultimately defining the boundaries within which patient agency is negotiated and realized.

\begin{table*}[!ht]
  \centering
  \caption{Demographics of our participants. In the patient block, the ``Paired Care Worker'' column lists the code of the patient’s matched care worker when that worker also took part in the study (P1$\leftrightarrow$CW7, P3$\leftrightarrow$CW4, P4$\leftrightarrow$CW8); a dash (-) indicates that the patient’s paired care worker did not participate. The lower block reports the occupations of HP and CW participants (doctor, nurse, nurse assistant, social worker, personal care aide). Aliases are unique, de-identified labels and do not imply ordering or hierarchy.}
  \label{tab:participants}
  \setlength{\tabcolsep}{6pt}
  \begin{tabular*}{\textwidth}{@{\extracolsep{\fill}} c c c c c}
    \toprule
    Alias & Age & Gender & Paired Care Worker \\
    \midrule
    P1  & 62 & M & CW7 \\
    P2  & 71 & M & - \\
    P3  & 67 & F & CW4 \\
    P4  & 58 & F & CW8 \\
    P5  & 81 & F & - \\
    P6  & 92 & M & - \\
    P7  & 85 & F & - \\
    P8  & 66 & M & - \\
    \addlinespace[0.8ex]
    \toprule
    Alias & Age & Gender & Occupation \\
    \midrule
    HP1 & 62 & F & Doctor             \\
    HP2 & 30 & M & Doctor             \\
    HP3 & 64 & F & Nurse              \\
    HP4 & 52 & M & Doctor             \\
    HP5 & 48 & F & Nurse              \\
    HP6 & 50 & F & Nurse              \\
    HP7 & 46 & F & Nurse assistant    \\
    CW1 & 35 & F & Social worker         \\
    CW2 & 47 & F & Social worker         \\
    CW3 & 55 & F & Social worker         \\
    CW4 & 40 & F & Personal care aide    \\
    CW5 & 61 & F & Personal care aide    \\
    CW6 & 58 & F & Personal care aide    \\
    CW7 & 54 & F & Personal care aide    \\
    CW8 & 47 & F & Personal care aide    \\
    \bottomrule
  \end{tabular*}
\end{table*}

\section{Methods}
Rather than analyzing the practices of a single stakeholder group, this study examined how patient agency is enacted, expressed, and represented among multiple stakeholders within care networks. To achieve this comprehensive understanding of HBC, we conducted a field study that combined interviews with ethnographic observation. The following sections describe participant recruitment, procedures of data collection and analysis, and ethical considerations throughout the research process.

\subsection{Participant}
To reflect the complex dynamics of HBC, we sought to capture diverse stakeholder perspectives within HBC networks. We recruited patients, healthcare professionals (doctors, nurses, and nurse assistants), and care workers (social workers and personal care aides). We deliberately excluded informal caregivers such as family members, neighbors, or friends to focus specifically on the roles and interactions of professional care providers within the HBC network.

Given the vulnerable nature of HBC patient populations, our recruitment approach prioritized both authentic data collection and participant safety. The primary inclusion criterion across all stakeholder groups was receiving or providing HBC services for at least three months at study initiation, ensuring sufficient experience with HBC dynamics. For patients, we established additional criteria to ensure safety: 1) ability to communicate effectively, 2) capacity to provide informed consent, and 3) absence of acute medical instability. 

We employed snowball sampling \cite{naderifar2017snowball} to identify care networks where stable, trusting relationships had already been established between patients and service providers. This approach was particularly important for observing authentic interactions without disrupting existing care dynamics, while also aligning with ethical principles for research with vulnerable populations by preventing potential exploitation through the research process \cite{jokinen2002ethical, cotton2021ethical}. We began by establishing contact with three medical institutions and seven care agencies in Seoul, through which we identified care networks characterized by ongoing, stable relationships. Within these networks, we then invited stakeholders who expressed interest in participation and met our inclusion criteria.

This recruitment process resulted in eight patients (P1-8), seven healthcare professionals (HP1-7), and eight care workers (CW1-8). All patients were under the care of HP4-7. Table \ref{tab:participants} presents participant demographics and their assigned codes, with paired relationships between patients and care workers indicated where applicable.

The average age of the patient participants was 72.75 years (range: 58-92, SD=11.97). Four patients (50\%) identified as male, and four (50\%) identified as female. Seven of the patients (87.5\%) were living alone. The average age of the healthcare professional participants was 50.29 years (range: 30-64, SD=11.28), and this group included three doctors, three nurses, and one nurse assistant. Among them, five (71\%) identified as female and two (29\%) identified as male. The average age of the care worker participants was 49.63 years (range: 35-61, SD=9.00), consisting of three social workers and five personal care aides, and all of them identified as female. One of these participants, CW3, was dually licensed as a social worker and a nurse but was working as a social worker at the time of the study. All participants were Korean. For compensation, we provided 70,000 KRW (approx. 50 USD) for interview participation and 40,000 KRW (approx. 30 USD) for observation participation.

\subsection{Data Collection}
We conducted a field study consisting of 23 interviews and 60 hours of ethnographic observation in parallel, aiming to connect experiences across stakeholder groups within HBC networks. Below, we describe how we conducted the interviews and observations.

\subsubsection{Interview study}
We conducted semi-structured interviews to explore diverse perspectives and experiences within the HBC network. To reduce self-censorship caused by group dynamics, interviews were conducted individually \cite{hollander2004social}. We developed three distinct protocols to reflect the context and expertise of each stakeholder group.

Patient interviews were conducted in their homes for approximately 30 minutes to minimize physical and cognitive burden on participants \cite{stevens2023best, samsi2020interviewing, hsu2025designing}. If a patient expressed fatigue or discomfort, the interview was immediately stopped and could be resumed at a later time \cite{samsi2020interviewing}. Before each interview, we obtained information from healthcare professionals and care workers about patients' health status and potentially sensitive topics, such as family relationships and functional abilities. We used this information to adjust our questions accordingly, avoiding unnecessary emotional distress to patients.

We used open-ended questions to elicit responses in patients' own words. When patients struggled to answer, we rephrased questions in simpler terms to reduce cognitive burden \cite{persson2023daily}. Our questions explored their daily priorities and values, perceptions of their current health status, experiences and preferences regarding care services, and participation in decision-making processes.

Interviews with healthcare professionals lasted approximately one hour and were conducted in person or remotely through Zoom. The interview protocol explored how the unique context of HBC shaped clinical practice and relationships with other stakeholders. We asked about differences between hospital-based care and HBC, the influence of home settings on clinical decision-making, priority-setting under constrained medical resources, collaboration within multidisciplinary teams, and the collection, interpretation, and documentation of patient information.

Interviews with care workers also lasted approximately one hour and were conducted in person or remotely (via Zoom or telephone). We asked about daily care routines and relationships with patients, information sharing with healthcare professionals, and perceptions of care labor. For both healthcare professionals and care workers, we encouraged them to illustrate concrete examples to ensure responses were specific and contextualized.

All interviews were audio-recorded and transcribed after receiving consent from participants. During transcription, participants were anonymized with assigned codes and any potentially identifying information was removed.

\subsubsection{Ethnographic Observation}

Our ethnographic observation was conducted to complement the patterns not fully revealed during the interview process and to explore how participants' self-reports aligned with their HBC practices. We additionally recruited participants for the observation from our pool of interview participants, observing their actual HBC environments for approximately 60 hours. To protect patients' living environments and privacy, we only visited their homes during regular visits of healthcare professionals or social workers \cite{locher2006ethical}.

\begin{figure}[!ht]
\centerline{\includegraphics[width=0.85\columnwidth]{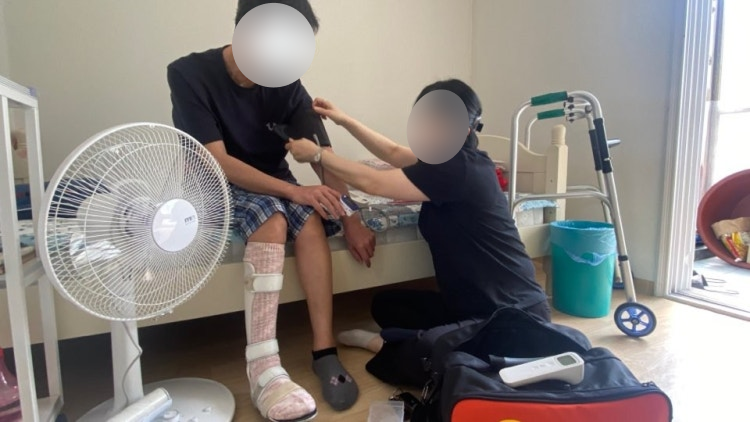}}
    \caption{HBC visit captured during our ethnographic observations with HP5 (right, seated).}
    \label{fig:hbc} 
    \Description{Home-based care visit. Nurse (HP5), seated on the floor, measures a patient’s blood pressure while the patient sits on a bed; kit, thermometer, walker, and fan nearby. Faces blurred.}
\end{figure}

Our observation of healthcare professionals involved accompanying them throughout the entire HBC process. This began with organizing patient information at the medical facility, followed by travel to patients' homes for in-home medical consultations, and concluded with the subsequent documentation of patient records. We also attended healthcare team meetings to observe the nature and scope of information being shared. Similarly, our observation of care workers covered their entire workflow, from departure to patients' homes, and to the process of writing and organizing documents. Additionally, we attended care worker meetings to examine their information-sharing practices. We conducted informal interviews before, during, and after observations to gain a deeper understanding of participants' perceptions, emotions, and thoughts. Information gathered and organized during observation sessions was written in field notes.

Concurrently, we collected patient-related documents created by healthcare professionals and care workers. We only reviewed records after receiving consent from both patients and the author of the record. We transcribed these documents on-site, excluding all personally identifiable information. During this process, we also conducted informal interviews about the intent behind the document and the challenges faced during documentation. In total, we collected 19 ``care reports'' written by care workers and 20 Electronic Medical Records (EMRs) created by healthcare professionals.

\subsection{Data Analysis}
Throughout the interview and observation, the research team held weekly meetings to discuss preliminary findings and identify themes within the data. After data collection was complete, the first two authors individually coded the interview transcripts and field notes using the open coding method from grounded theory \cite{corbin1990grounded}. All authors then iteratively reviewed and refined these codes in the weekly meetings to finalize the key findings. We also conducted informal checking with participants on occasion, to ensure the perspectives of researchers were not biased and to reintegrate feedback from the field into our themes.

The themes derived from this analysis offered insights into how the agency of HBC patients is expressed, and the factors within HBC that broaden the representation gap of this agency to HBC networks. Ultimately, by comparing and discussing these insights with prior research, we developed design considerations for effectively integrating patient agency in the HBC environment.

\subsection{Ethical Considerations}
Our study was approved by the Institutional Review Board of KAIST. All authors conduct research at the intersection of HCI and healthcare and have extensive experience working with vulnerable populations. Notably, one of the co-first authors is an M.D. who was practicing as an HBC doctor during the study. While this author's own clinical activities were not recorded as formal observational data, their personal experience enhanced the contextual understanding during data analysis, similar to how Kientz and Abowd became part of a therapist team to directly understand information needs for system design \cite{kientz2008designer}.

All research protocols were designed in reference to prior work on methodologies for qualitative research with patients having diminished cognitive and physical abilities and in their living environments \cite{hsu2025designing, houben2024design, yaron2024constant}. The protocols were then reviewed by two HBC doctors, including the aforementioned author.

Informed consent was obtained from all participants following a thorough explanation of the study's purpose, procedures, and methods. For patients in particular, we sought their consent before every observation session to ensure they maintained control over the process \cite{evans2020processes}. We guaranteed strict confidentiality, assuring participants that any issues raised would not be disclosed. Additionally, explicit consent was obtained before each session for any audio recording or photography. Prior to the study, we informed participants of their right to request modifications to the research protocols if they felt uncomfortable and to withdraw at any time without penalty. No participants requested modifications to the research process, and none withdrew from the study.

\section{Findings}
The HBC environments we observed provided integrated care to patients with multiple chronic conditions or mobility limitations through 3-6 hours of daily care worker support and monthly or biweekly visits from healthcare professionals. These conditions appeared conducive to patients exercising agency over the care process. Indeed, most service providers acknowledged the principle that the ultimate goal of HBC is providing care that respects patients' agency and choices. HP1 reinforced this commitment: ``\textit{PCC is important. We therefore try to incorporate patients' requests into our care approach.}'' Indeed, we observed multiple efforts to operationalize patients' expressed agency in practice. CW1 shared a concrete example: ``\textit{We sometimes adjust our care plans to enable patients to engage more frequently in activities they enjoy.}''

However, our findings reveal a systematic disconnect between patients' expressions of agency and the integration of this agency into HBC. To understand this complex phenomenon, we organize our findings around three interconnected themes. First, we examine the distinctive manifestations of patient agency within HBC environments and how service providers recognize and respond to these expressions (Section~\ref{subsec:4-1}). Next, we analyze barriers preventing the integration of patient agency into HBC across information practices (Section~\ref{subsec:information-practices}) and relational dynamics (Section~\ref{subsec:relational-dynamics}).

\subsection{Patient Agency and Its Recognition in HBC Networks} \label{subsec:4-1}
In this section, we examine how patient agency manifests in HBC across three distinct dimensions: maintaining continuity in everyday life, fostering self-efficacy through micro-achievements, and deepening care relationships through contextual understanding.

\subsubsection{Maintaining Continuity in Everyday Life} \label{subsubsec:4-1-1}
HBC patients prioritized maintaining continuity in their daily routines over pursuing future-oriented goals. 
P1 articulated their present-focused orientation: ``\textit{Thinking about the future makes me more anxious. It's more comforting to think about how to spend each day at home.}'' This orientation sometimes manifested as resistance to institutional care services. CW1 stated, ``\textit{Patients refuse to go to a nursing home, even when their physical condition necessitates it. They insist they would rather die at home.}'' This represented not mere stubbornness but a powerful assertion of agency aimed at preserving their life context. Notably, even HBC services delivered in patients' familiar home environments could be perceived as intrusive. CW3 noted, ``\textit{Some patients expressed discomfort with ``strangers'' entering their personal space to discuss care services.}''

Recognizing these characteristics, service providers began reconsidering traditional approaches to medical service delivery. Rather than unilaterally providing medical and care services, they adopted an approach that prioritized listening to patients' stories and understanding their requests, and they had to establish trust relationships from their position as guests. HP2 emphasized the importance of this approach: ``\textit{HBC represents a completely different environment where doctors' expertise and authority are not readily established. Without emotionally attuned conversations, building rapport becomes extremely difficult.}''

\subsubsection{Fostering Self-Efficacy Through Micro Achievements} \label{subsubsec:4-1-2}
Beyond the desire to maintain continuity in everyday life, HBC patients demonstrated another form of agency through their efforts to recognize and maintain their functional capabilities within daily routines. Interestingly, their goals focused not on active recovery or social reintegration, but rather on maintaining basic functions such as Activities of Daily Living (ADL) and Instrumental Activities of Daily Living (IADL) \cite{katz1983assessing, bae2022activities}.
P3 expressed, ``\textit{If I could walk well on my own, I'd be able to hang up the laundry and do my own cleaning, but I can't, so it's really frustrating,}'' revealing a strong aspiration to perform daily activities independently. 
Similarly, P4 and P5 both expressed their hope of basic physical function recovery, stating, ``\textit{I want to stand up.} (P4)'' and ``\textit{I just want to be able to sit up.} (P5)'' 
P1 defined their emotional well-being through small daily activities such as ``\textit{watching comedians on YouTube and staying cheerful.}'' These examples demonstrated that HBC patients underscored maintaining and recovering micro-functions of daily life rather than pursuing ambitious treatment goals.

More importantly, this perception was not formed in isolation but was co-constructed through interactions with care networks. HBC patients recognized subtle changes or improvements that they might not have noticed, through the observations and feedback of service providers. HP2 explained this dynamic: ``\textit{Patients value moments when they notice their appetite returning or regain some movement after complete immobility. They seem happiest when doctors inform them that their blood test results have improved.}''

This co-construction process was bidirectional. Service \allowbreak providers also experienced fulfillment by witnessing and sharing in patients' small changes and achievements. HP4 reflected, ``\textit{I felt a sense of accomplishment when patients who used to swear and react harshly began to express gratitude and welcome us over time.}'' This suggests that patients' achievements in the HBC environment were not merely personal experiences, but relational phenomena that were mutually recognized and reinforced within care relationships.

\subsubsection{Deepening Care Relationships Through Contextual Understanding} \label{subsubsec:4-1-3}
The regular and repeated HBC enabled relationships between patients and service providers that went beyond simple care service delivery, and the depth of these relationships became a foundation for both the expression and recognition of patient agency. 

Our study revealed that service providers sought to develop relationships with patients in a staged manner. Initially, they focused on identifying medical needs and health requirements. CW3 described this process specifically: ``\textit{When we go out for initial consultations, we conduct needs assessments for each patient across multiple categories.}'' Over time, however, they came to understand the contexts of patients' lives and personal characteristics. HP5 described this experience through the case of a patient with hemiplegia from stroke. Despite the mobility difficulties, they tried to maintain consistent exercise routines. HP5 learned from photos on the wall that the patient had been an athlete, which provided a new perspective for understanding their persistent desire for exercise. 

This example illustrates the unique strength of the HBC environment that extends beyond conventional medical information gathering. Patients' homes served as repositories of contextual information that revealed their life histories, preferences, and values. Everyday elements such as furniture arrangements, food in refrigerators, photos on walls, and bed positions provided clues for developing a deeper understanding of patients. HP1 also recognized the importance of such contextual information, stating, ``\textit{We sometimes photograph and record the condition of patients' blankets or beds.}''

The depth of relationships was further strengthened when service providers and patients discovered and shared common interests. CW7 described the relationship with a patient as follows: ``\textit{The patient used to raise dogs and loves talking about them. Their care worker also raises a dog, so they have many conversations about this common interest.}''

These relationships often transformed into specific, personalized care plans. For example, HP7 described attempting creative interventions based on patients' contextual information, such as ``\textit{reading fairy tales to comfort a patient with low cognitive function}'' or ``\textit{inquiring about pets' well-being with a patient who has dementia.}'' Thus, the relationships formed in HBC environments provided a foundation where patient agency could be understood and respected within the broader context of their lives.

\subsection{Information Practices to Integrating Patient Agency in HBC} \label{subsec:information-practices}

\begin{figure*}[!ht]
\centerline{\includegraphics[width=0.8\textwidth]{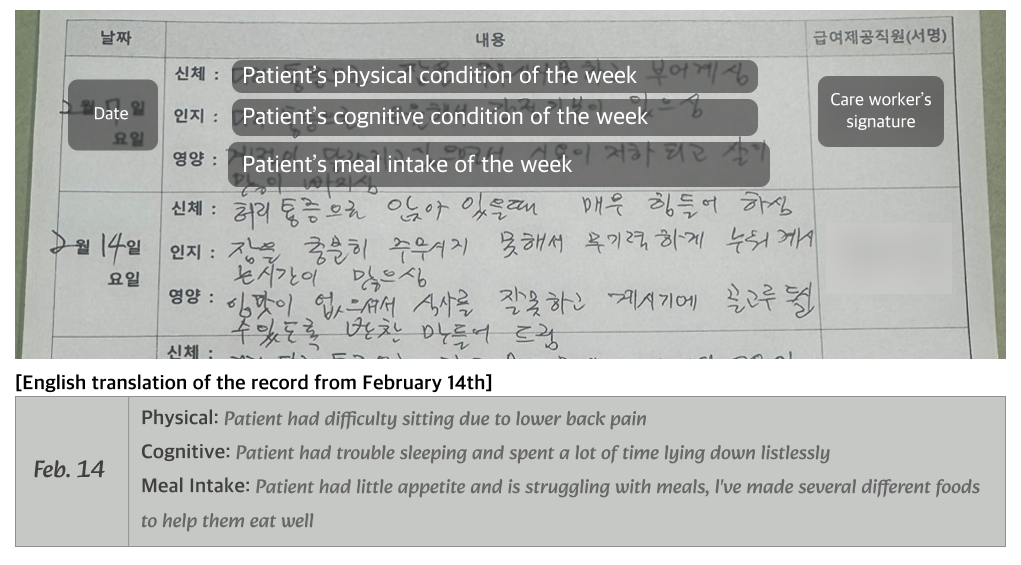}}
    \caption{A weekly care report captured during our observations with the care agency where CW1 works. The top panel depicts the original handwritten record in Korean, with overlays that label the core schema. The bottom panel presents our English translation of the entry from February 14th.}
    \label{fig:reports} 
    \Description{The top panel shows the original Korean report with overlays labeling key fields: date; patient’s weekly physical condition; cognitive condition; meal intake; and the care worker’s signature area. The bottom panel shows the English translation of the same record dated February 14. Physical: Patient had difficulty sitting due to lower back pain; Cognitive: Patient had trouble sleeping and spent a lot of time lying down listlessly; Meal intake: Patient had little appetite; several foods were offered to help them eat well. All identifying information is anonymized.}
\end{figure*}

As demonstrated in Section \ref{subsec:4-1}, HBC environments provided rich opportunities for unique expressions of patient agency, and service providers showed willingness to respect these expressions. However, we also observed systematic disconnects that occurred when translating these positive perceptions and intentions into actual shared care plans.
Therefore, in this section, we examine the information practices in HBC to identify the causes of such disconnects, organizing our findings by following the key stages of the practices.

\subsubsection{Provider Preconceptions and Observational Constraints in Information Acquisition} \label{subsubsec:acquisition}
We identified that service providers often failed to recognize patient agency during the process of acquiring information. The cognitive and physical limitations of HBC patients, combined with their low health literacy, led service providers to perceive them as fundamentally passive. CW1 stated, ``\textit{Patients often can't understand the terminology itself, so they couldn't explain what kind of disease they have.}'' HP2 similarly questioned patients' expressive capabilities: ``\textit{They are generally very passive in most cases, since they often can't articulate that they're in pain.}'' 

This preconception resulted in the systematic exclusion of patient agency from this stage. 
CW2 shared their experience: ``\textit{Since the patients can't really understand what's going on, doctors often end up talking to the care workers instead of directly to the patients. Some patients picked up on this and felt uncomfortable about it, like they were being treated as invisible.}'' This expression of discomfort about being treated as ``invisible'' was a clear demonstration of patient agency—their capacity to perceive situations and express. However, even these signals were not properly recognized due to the preconceptions of patients.

The structural characteristics of HBC services—intermittent observation—also imposed significant constraints on this stage. Intermittent visits made it difficult for care providers to identify gradual changes or subtle patterns of patients. CW1 stated, ``\textit{Patients' mood and condition vary each day, so we can't assess the overall situation just from our few hours of observations.}'' 

Such snapshot-style observations risked reducing the dynamic complexity of patients' daily lives to fragmentary and static images. The complexity of this problem was clearly illustrated in a case shared by CW4: ``\textit{One patient said to the care worker that they've got no appetite. But we eventually discovered that when they're alone, they eat really well.}'' In this case, the patient exhibited a different behavioral pattern when with service providers versus when alone. However, it was difficult to determine whether this difference stemmed from the patient's intentional choices (e.g., exaggeration to receive help), relational influences, or other factors.



Consequently, intermittent observation created systematic misunderstandings about patient agency. Service providers tended to generalize their observations from limited periods as representing patients' overall condition, leading them to perceive patients as more dependent and passive than they actually were. This misunderstanding created a vicious cycle in which patients' spontaneous attempts or choices were not reflected in shared care plans.


\subsubsection{Structured Documentation Systems Filtering Out Contextual Understandings} \label{subsubsec:documentation}

The first barrier in information documentation was that structured documentation systems used by each profession failed to capture rich expressions of patient agency. Each service provider in HBC networks used structured documentation forms for efficiency and medical necessity. Healthcare professionals recorded standardized medical information such as medical history, chief complaints, current medications, and vital signs through EMRs. Care workers used care reports to document key patient information, including their physical and cognitive conditions with meal intake (See Figure \ref{fig:reports}). Care reports were written daily and weekly.

While these structured forms appeared systematic and comprehensive, they failed to capture key evidence of patient agency in the HBC environment. Notably, these forms commonly lacked space to record contextual information and failed to provide systematic guidelines. 
CW6 articulated this challenge: ``\textit{It's really hard to recall and document what I did. There's a sense of being overwhelmed about what to write.}'' This was not simply a matter of recording ability. The subtle emotional changes, activity preferences, and values revealed through personal histories that service providers observed did not fit within existing form categories. 

Furthermore, disparate recording systems intensified information gaps between stakeholders. HP2 clearly identified this problem: ``\textit{Welfare centers know well about welfare environments, but may not know much about medical situations. In contrast, medical institutions tend to prioritize medical conditions and place sociocultural backgrounds in lower priority.}'' 

Consequently, structured recording systems provided efficiency while reducing diverse expressions of patient agency to standardized categories. Patient-initiated preferences, refusals, and suggestions were briefly noted in `other' sections or omitted entirely, leading to systematic exclusion of patient agency when establishing shared care plans.

\subsubsection{Informal Communication Channels Distorting Patient Voice}
Recognizing the limitations of documentation systems, service providers attempted to bridge information gaps through information sharing. However, due to the absence of formal communication channels, they relied on informal, unsystematic channels such as group chats, phone calls, or meetings. By their very nature, these channels lack the official structure, record-keeping, and clear accountability required in formal settings. As a result, participation becomes arbitrary, leading to the frequent exclusion of crucial stakeholders such as care workers or doctors. CW3 explained this structural exclusion: ``\textit{There's a group chat with everyone from the doctors to nurses, but not all care workers are included.}''

Information sharing through such informal channels resulted in inconsistent interpretation of information. Each stakeholder interpreted limited information solely from their perspective, leading to a fragmented understanding of patients. CW2 recalled, ``\textit{There was a case where doctors had different opinions about a patient. We were confused about which opinion we should focus on.}'' This absence of formal communication could threaten patient safety. Another case shared by CW3 illustrated this risk: ``\textit{There was a patient who continuously had trouble eating. Their care worker just thought it was because of their age, but then they suddenly collapsed due to constipation. (\dots) If we'd had a way to share this information quickly, this whole thing probably could've been avoided.}'' 

\subsubsection{Fixed Narratives Stigmatizing Patients}
Due to the long-term nature of HBC, information continuously accumulates. However, this paradoxically resulted in forming fixed narratives that stigmatized patients and overlooked their expressions of agency. For example, when a patient was initially recorded as ``uncooperative'' or ``passive,'' such labels were repeatedly transmitted through documentation. CW5 identified this problem: ``\textit{If we hear `this patient is difficult,' we tend to perceive even normal requests from them as being difficult.}''

These fixed narratives blocked recognition of patient change. Even when patients became more cooperative over time, demonstrated new interests, or expressed different preferences, these changes were buried under existing narratives and received less attention. Instead, service providers tended to interpret patients' new attempts as ``unusual behavior.'' A case shared by HP5 demonstrated how these stereotypes create misunderstandings and conflicts. When a patient experiencing depression retrieved a long rope from their room, a care worker interpreted this as a suicide attempt based on the accumulated information. However, the patient had actually retrieved it to organize their clothes. This case illustrated how accumulated information can cause current situations to be viewed through distorted lenses. 

Consequently, the accumulation and reproduction of information objectified patients as beings with fixed characteristics rather than dynamic and changing subjects, causing shared care plans to be based on past stereotypes rather than patients' current needs and preferences.

\subsection{Relational Dynamics in HBC Networks} \label{subsec:relational-dynamics}
In the previous section, we examined how patient agency is diminished or distorted within information practices. During our observations, we found that these problems extend beyond technical or procedural issues. The production and interpretation of information are inherently social processes, deeply shaped by relationships among stakeholders, hierarchies of expertise, and levels of mutual trust. Therefore, this section explores how these relational dynamics further amplify the representation gap of patient agency.

\subsubsection{Dilemmas Faced by Patients When Asserting Their Agency} \label{subsubsec:dilemmas}
As discussed earlier, patients want their agency integrated into shared care plans, but this requires support from the care networks. 
This tension between agency and dependence manifests in two paradoxical situations within relationships. First, close ties with service providers can become barriers to requesting legitimate help. Patients worry that asking for assistance undermines their agency by acknowledging vulnerability or imposes a burden on service providers. CW1 illustrated, ``\textit{There was this patient I'd been taking care of for a while. Their condition got worse, and they couldn't get to the toilet on their own. They had to call me whenever they needed to go, and it was clearly taking a toll on them mentally. I was worried it might damage their sense of dignity.}''

Second, because care unfolds in the home—the most intimate and private space for patients—professional boundaries blur and sometimes generate conflict. Some patients treat care workers or doctors as personal helpers rather than professional providers, leading to disrespectful behavior. HP3 noted, ``\textit{When we meet family members who treat care workers as if they are in a master-servant relationship, it becomes extremely difficult.}'' CW3 added, ``\textit{Obviously, there are cases of verbal abuse or disrespect in HBC.}''

\subsubsection{Invisible Labor and Role Conflict of Care Workers}
Alongside the patients’ dilemmas, the role conflict experienced by care workers constitutes another relational barrier to reflecting patient agency within HBC. In many countries, care workers are certified professionals \cite{kash2007community, kim_occupational_2020}. Yet they often confront expectations from patients to serve primarily as housekeepers, performing cooking and cleaning. CW1 observed, ``\textit{Patients want better services, but from the care worker's perspective, those requests often feel like they're being asked to do basic housework.}'' As a result, the professional aspects of care work become invisible, overshadowed by visible domestic tasks. 

Many care workers enter the field to promote health, but workplace conditions that underutilize their clinical skills create role conflict. According to CW4, care workers say, ``\textit{I came to support patients’ well-being, but I end up doing housework.}'' This erodes motivation and contributes to de-skilling. CW6 described this tension, stating ``\textit{Some care workers only want to do housework and cleaning. When they're asked for their education, they sometimes refuse.}''


\subsubsection{Reproduction of Doctor-Centered Relational Dynamics} \label{subsubsec:doctordynamics}
To integrate patient agency into shared care plans, medical indicators must be considered alongside patients' everyday contexts and relational needs. Despite HBC occurring in patients' homes, decision-making remains centered on doctors, thereby reproducing structural dynamics that render patients' life contexts and preferences secondary.

In a case illustrated by HP4, a patient in their seventies with diabetes, hypertension, and mild dementia developed a pressure ulcer after spending all day sitting on the floor with their dog. HP4 recommended switching to a bed and using cushions to prevent pain and ulcers. The patient, however, strongly preferred floor living with their dog and only wanted pharmacological pain management. This case demonstrates that when patients' preferences are not legitimized alongside clinical reasoning, patients are pushed into defensive postures, retreating to narrow forms of participation focused on requesting medications.

This hierarchy of information creates a paradoxical positioning. Although patients are experts on their own lives, they are located as ``non-experts'' in medical decision-making. HP1 candidly reflected, ``\textit{I’m not sure we paid much attention to what the patient proactively proposed. I think we regarded those as unimportant.''}

This also discourages patients from expressing experiential knowledge. Patients learn to communicate only in a symptom legible to doctors, narrowing their holistic needs. HP6 shared the case of a patient in their sixties undergoing dialysis with diabetes, hypotension, and chronic kidney disease. The team responded primarily with medication adjustments and routine monitoring. Eventually, the patient concluded that the service no longer added value to their life and requested discontinuation of HBC. This case illustrates how patients become fixed as passive recipients when care plans rely solely on the doctor's judgment.

The less patient agency is reflected in shared care plans, the more passive patients become, deepening the hierarchy of information in a vicious cycle. Ultimately, the reproduction of doctor-centered dynamics is not merely a matter of attitudes of individual providers; it stems from structural features all designed around biomedical outcomes.

\section{Discussion}
In this section, we discuss the nature of patient agency in HBC as a dynamic capacity shaped by relationships and the material home environment, rather than a static individual attribute. We then examine how current sociotechnical infrastructures often fail to capture these nuances, resulting in a representation gap that obscures patient inputs in shared care planning. Based on these insights, we propose design considerations for technology to bridge this gap and better integrate patient agency into the coordination of HBC.

\subsection{Understanding Patient Agency in HBC Through Care Relationships} \label{subsec:agency}
Our study illuminates that patient agency in HBC is not a static attribute of the individual but a dynamic capacity negotiated through the continuity of daily life, recognition from others, and engagement with the material home environment. 
By shifting focus from individual capability to these relational dynamics, we reveal the mechanisms through which agency is realized and overlooked within the constraints of HBC.

Specifically, our findings challenge the prevailing perception observed in HBC networks that physical dependence equates to a lack of agency. As noted in Section~\ref{subsubsec:acquisition}, we found that service providers often perceived patients as ``fundamentally passive,'' implicitly assuming that the inability to execute tasks independently signaled an inability to make meaningful decisions. However, our analysis reveals that patients exercised agency not by rejecting support, but by asserting their identity to \textbf{preserve a sense of self within dependence}. The case shared by HP4, the patient who persisted in sleeping on the floor with their companion dog, illustrates this vividly. While a clinical perspective might interpret this refusal of a bed as non-compliance, it was in fact a deliberate effort to protect a cherished routine and emotional connection with their own environment. This indicates that agency in HBC operates not through achieving physical independence, but through sustaining the meaning of one's life in their dependence within relationships and surroundings \cite{sherwin1998politics, lonkila2021care}.

Furthermore, this study extends the understanding of patient agency in HBC as an intersubjective phenomenon. We observed that a patient's sense of efficacy was not achieved in isolation but was co-constructed through the gaze of the care provider. Patients often gauged their condition not through standardized metrics but through the provider's validation of achievements.

This finding aligns with the perspectives of care ethics \cite{tronto2020moral}, while adding practical insights. In HBC, the provider's role goes beyond mere support. It involves recognizing agency expressed through small acts that invite acknowledgement from others, such as thanking caregivers, sharing stories, or politely refusing help. These micro-interactions allow the care networks to see patients as active moral agents rather than passive recipients of care. Without this \textbf{mutual recognition}, in which patient agency is interpretively and performatively constructed, the patient's internal experience of agency remains invisible and unverified \cite{acke2022one}.

Our data also highlights the active role of \textbf{the material home environment in mediating agency}. Consistent with the perspective of recognizing patient agency as a co-constructed trait through relationships with human and non-human stakeholders, we found that the home was not merely a backdrop but an active participant in the care network \cite{mol2008logic}. Photographs, furniture arrangements, and personal objects functioned as `boundary objects' that allowed providers to see the patient not just as a medical case, but as a person with a history \cite{reddy2001coordinating}. For instance, as shown in Section~\ref{subsubsec:4-1-3}, discovering a patient's past as an athlete through photos allowed HP5 to reframe the patient's desire for exercise. This suggests that patient agency in HBC is distributed across the home's material environments, and that preventing its erasure requires the care network to actively engage with these environments.

Yet, we found that \textbf{intimate relationships sometimes created tensions} that could paradoxically constrain agency. As the boundaries between professional service and personal bonds blurred, patients often hesitated to assert their needs to avoid burdening the provider or disrupting the relationship. An example is CW1's account of a patient who felt deep distress when requesting toileting assistance, fearing that exposing such vulnerability would compromise their dignity in front of a care worker they had come to value personally. This reveals that the interdependencies which support patient agency can also impose emotional constraints, requiring patients to navigate the delicate tension between accepting help and preserving their dignity within the relationship \cite{cournoyer2021conditional, roth2009emotional}.

\subsection{Understanding the Representation Gap in Integrating Patient Agency into HBC} \label{subsec:role}
The findings of this study reveal a systematic representation gap where the expressions of patient agency fail to be integrated into shared HBC plans. Here, we argue that this gap is not merely a product of individual negligence or role ambiguity, but is actively produced by the sociotechnical infrastructures of HBC that fragment care into disconnected domains \cite{fitzpatrick2013review}.

Our analysis indicates that the institutional logics embedded in \textbf{current documentation artifacts systematically filter out expressions of patient agency}. Specifically, the structured forms used by healthcare professionals (EMRs) and care workers (Care Reports) are designed primarily for medical safety and administrative accountability. Consequently, these artifacts function as filters of patient agency, rendering the relational and contextual aspects of agency invisible before they can enter the deliberation process~\cite{pine2014}.

Moreover, this representation gap is exacerbated by \textbf{the absence of sociotechnical artifacts capable of bridging heterogeneous perspectives}. In distributed care networks, effective coordination typically relies on boundary objects that allow different communities to align their work without compromising their specific viewpoints \cite{lee2006human}. However, our findings show a critical lack of such coordinative objects in HBC. Healthcare professionals and care workers operate in parallel information worlds, relying on informal channels like group chats to bridge this gap. As noted in Section~\ref{subsubsec:dilemmas}, the lack of a common information space means that critical insights—such as the discrepancy between a patient's public behavior and private capabilities—remain isolated within specific dyads and are never synthesized into a holistic understanding. Without a shared infrastructure to negotiate these diverse interpretations, the patient's voice is fragmented and lost in transmission \cite{bardram2005}.

These \textbf{infrastructural limitations actively reproduce and solidify doctor-centered power dynamics}. As noted in Section~\ref{subsubsec:doctordynamics}, when conflicts arose between clinical indicators and patient preferences, the complex data of the medical record invariably overrode the knowledge of the patient's life context. By prioritizing biomedical metrics over experiential knowledge, the current HBC infrastructure structurally positions the patient as an object of observation rather than a subject of decision-making \cite{reddy2001coordinating}.

Therefore, the failure to integrate patient agency into HBC is a fundamental sociotechnical challenge. The current HBC infrastructures
do not merely fail to capture patient agency but actively reshape what counts as legitimate knowledge in care coordination, thereby suppressing the relational, contextual, and materially distributed forms of HBC patient agency. To bridge this representation gap, we have to move beyond simply encouraging stakeholders to listen better and instead support the artifacts and infrastructures to recognize and circulate the diverse forms of knowledge that constitute patient agency.

\subsection{Design Considerations to Integrate Patient Agency in HBC}
Drawing from our relational analysis of patient agency (Section~\ref{subsec:agency}) and the infrastructural barriers identified in shared care planning (Section~\ref{subsec:role}), this section proposes specific design considerations to integrate patient agency in shared HBC plans. These considerations position technology as infrastructure that mediates diverse knowledge forms, makes invisible patient agency visible, and bridges the representation gap between patient expressions and shared care plans.

\subsubsection{Embedding Patient Values to Configure the Boundaries of Dependence} 
Section~\ref{subsec:agency} established that agency in HBC is often expressed by preserving one's identity within the context of receiving care, but the institutional logic that prioritizes safety and efficiency overrides their personal way of being. Consequently, patients are reduced to passive recipients of care, lacking a mechanism to assert their own standards. Therefore, HBC technology should be designed specifically to \textbf{enable patients to define the boundaries of their dependence}.

To achieve this, technology should translate patient values into the guiding principles for care interventions \cite{mroz2023acceptability, ladin2023effectiveness}. In a state of physical dependence, a patient's values (e.g., dignity, continuity of habits, privacy) are easily compromised by the provider's routine. Bridging this gap requires translating these abstract values into concrete guidelines for shared care planning \cite{corbett2020planning}. Recent advances in artificial intelligence (AI) further support this by synthesizing heterogeneous data sources for healthcare decision support \cite{alhejaily2024artificial}.

For example, consider a patient who values the companionship of a pet over standard clinical hygiene (as in the case shared by HP4). Rather than simply enforcing a switch to a bed, an AI-enabled system could interpret this preference as a guiding principle derived from the patient's values. It could then generate alternative safety measures that align with both safety and the patient's values, such as \textit{recommending pressure-relieving floor mats}.

By \textbf{treating patient values as rigid design parameters} that the care plan must accommodate, technology shifts the nature of dependence. Instead of unconditionally surrendering control to the caregiver, the patient can specify how assistance should be rendered. This ensures that while the execution of tasks remains in the hands of others, the manner of care aligns with the patient's will rather than an imposition of external standards.

\subsubsection{Facilitating Mutual Recognition Through Feedback Loops}
Section~\ref{subsec:agency} established that patient agency is an intersubjective phenomenon requiring mutual recognition, where patients gauge their efficacy through providers' validation of achievements. However, Section~\ref{subsec:role} demonstrated that documentation systems function as filters that privilege biomedical metrics over the contextual narratives where patient achievements reside. 
To address this, technologies in HBC should be designed to \textbf{facilitate a continuous feedback loop} in care networks, by recording and responding to patient-reported achievements. 

Prior work suggests that AI and multimodal sensing can serve as effective mediators of user self-expression and disclosure \cite{kim_mindfuldiary_2024, kocielnik2018reflection, nepal2024mindscape}. These affordances allow patients to record everyday achievements naturally using voice, images, or sensors \cite{gomez2023design}. Specifically, they can enable patients to express micro-achievements, experience a sense of accomplishment, and transmit these signals to the care network in structured ways.


For instance, consider a patient who manages to do their own laundry. Using a voice interface, the patient records a brief reflection: ``\textit{It was tiring, but I finished the laundry on my own today.}'' The interface processes this audio using natural language understanding to tag the event as a success in ``\textit{maintaining instrumental activities of daily living (IADL)}'' and forwards it to the healthcare provider. The provider, prompted by this update, replies with a message: ``\textit{I received your update. Handling laundry is a big task. It shows your dexterity is holding up.}'' The interface then allows the patient to acknowledge this feedback by sending a simple voice note of gratitude or a reaction. This sustains the feedback loop, rather than a one-off transaction.

By making patient contributions visible and ensuring a reciprocal response, this infrastructure concretizes the mutual aspect of care. Moreover, it transforms documentation from a surveillance tool into a channel for continuous engagement, challenging the logic that currently renders micro-achievements invisible \cite{nie2024developing, kononova2019use}.

\subsubsection{Leveraging Material Home Environment to Construct Boundary Objects}
Section~\ref{subsec:agency} demonstrated that material home environments actively mediate agency, while Section~\ref{subsec:role} identified the critical absence of boundary objects that enable stakeholders to align on interpretations of the rich contextual information in these environments. Therefore, technologies in HBC should be designed to \textbf{construct boundary objects by leveraging contextual information from the home environments} \cite{lee2006human}.


Since context can be gathered across modalities, Multimodal Large Language Models (MLLMs) can support the capture and integration of contextual richness from home visits (e.g., photographs, voice notes, documents) into coherent representations \cite{yin2024survey}. 
For instance, consider an MLLM-embedded documentation system capturing photos of a patient's past as an athlete, alongside a voice note expressing frustration with being bedbound. Instead of a standard care document, the system constructs a ``\textit{Training Schedule}'' as a shared care plan tailored to the patient. Within this plan, the patient's desire to exercise is reframed to align with their history rather than conflicting with care protocols: the doctor sets safety parameters (defining the \textit{training limits}), the care worker logs the assistance provided (monitoring \textit{form}), and the patient tracks progress toward a goal (recording \textit{stats}).

This ensures that, while each stakeholder applies their specific expertise, they collaborate on a single, meaningful artifact. By leveraging contextual information from the home environment, the technology creates a common information space where safety and identity are no longer opposing forces but integrated components of the patient's daily life.

\subsubsection{Scaffolding Reciprocity Through Reframing Requests into Collaborations}
Section~\ref{subsec:agency} identified that intimate relationships paradoxically constrain agency when patients hesitate to assert needs, fearing they will become a burden. This hesitation stems from framing care as a one-way transfer of service. To mitigate this, we need \textbf{communication protocols that transform requests for assistance into care collaborations}, and establish a reciprocal dynamic.

Specifically, HBC should restructure the ``help-seeking'' interaction. Rather than framing the patient as a passive recipient asking for a favor, the system should position them as an active manager of their care. This requires protocols where the patient not only initiates tasks but also validates their completion, turning the interaction into a bidirectional exchange of value.

Interfaces should be designed to assist patients in making requests through collaborative prompts, such as ``\textit{Let's address this together}'' rather than ``\textit{Request help}'' \cite{mohr2011supportive}. Furthermore, once the task is completed, the interface should prompt the patient and the care provider to offer small gestures and express gratitude \cite{day2020gratitude}.

By presenting care tasks as collaboration rather than a one-sided favor, and by embedding opportunities for mutual acknowledgement, the system creates a sense of shared accomplishment. This puts the relational aspect of agency into practice, ensuring that the act of receiving care becomes a reinforcement of connection rather than a source of shame \cite{mohr2011supportive}.

\subsubsection{Democratizing Information Infrastructures Through Attribution and Temporal Accountability} 
This study revealed that current documentation artifacts function as institutional filters privileging biomedical metrics, while informal communication channels fragment critical insights. Moreover, accumulated narratives stigmatize patients as fixed cases.

To redistribute the authority, documentation artifacts should be designed to \textbf{explicitly attribute each observation and interpretation to its source}. For example, documentation systems should display: ``\textit{The care worker noticed the patient was struggling to digest dinner. The doctor interpreted this as a sign of gastritis and recommended a hospital visit.}'' This makes visible whose knowledge shaped decisions and whose was excluded.

Additionally, structured reflection prompts can facilitate perspective-taking: ``\textit{What did you learn from the care worker's observation?}'' or ``\textit{How has this changed your understanding of the patient?}'' \cite{muller2024developing, yuan2025day}. These prompts help stakeholders reflect on their own roles while also appreciating the expertise of others.


To prevent stigmatization, \textbf{temporal framing should be adopted} to the design of documentation artifacts in HBC \cite{butler2021there}. Historical entries should carry explicit timestamps and circumstances: ``\textit{Resistance to medication \textit{(documented 3 months ago)}.}'' Recent changes should be emphasized with equal weight to older ones, while allowing patients to append updates. Such temporal framing helps position patients not as fixed by their past but as changing and evolving subjects \cite{mcadams2019first}.

\subsection{Limitations and Future Work}
In this section, we discuss the limitations of our study that could impact the generalization of the findings. First, there is a limitation regarding the selection criteria for patient participants. To vividly capture the dynamics in the field, we adopted the capabilities of verbal communication and expression as one of the inclusion criteria for patient recruitment. However, this approach does not fully represent the spectrum of HBC patients, since many of whom have significant cognitive or physical impairments that lead to considerable communication difficulties \cite{guthrie2018, jespersen2025home}. Therefore, future research must explore alternative and non-verbal methods to understand how the agency of patients with limited communication abilities can be expressed and respected, ensuring their perspectives and intentions are not inadvertently excluded

Furthermore, this study was conducted within the specific regional and cultural context of Seoul, South Korea. However, HBC systems, healthcare policies, and cultural norms surrounding elder care can vary significantly across countries and regions \cite{llena2025countries}. Additionally, by using snowball sampling to recruit participants from networks with stable and trusting relationships, our study may not have fully captured the dynamics within less stable or conflict-ridden networks. Consequently, future work should include comparative studies across diverse cultural and healthcare systems to test the generalizability of our findings and to analyze the impact of contextual specificity on patient agency.

Finally, our research focused on exploring problems through fieldwork and proposing conceptual design considerations to address them. Future research should therefore involve developing systems based on the design considerations proposed in this paper and deploying them in real HBC settings. This would allow for long-term field studies with patients, healthcare professionals, and care workers to empirically evaluate whether the proposed designs can effectively bridge the representation gap in patient agency and improve the quality of care coordination.

\section{Conclusion}

Through 23 multi-stakeholder interviews and 60 hours of ethnographic observations, we examined how patient agency manifests in HBC settings and why it remains inadequately integrated into shared care plans. 
Our findings characterize agency in HBC not as a static individual trait, but as a relational capacity shaped through maintaining everyday continuity, validating micro-achievements, and navigating interdependent care relationships. However, we identified a significant representation gap, in which these subtle practices are often rendered invisible by rigid documentation infrastructures and hierarchical information practices that privilege biomedical metrics over contextual life-world knowledge. Drawing from these insights, we propose design considerations to configure the boundaries of dependence, facilitate mutual recognition and reciprocity, construct context-aware boundary objects, and democratize information infrastructures.
This work contributes to HCI by revealing that patient agency in HBC requires different conceptual and technological approaches to surface and integrate the subtle ways patients exercise agency within their home environments.


\begin{acks}
We thank our study participants and the reviewers for their valuable feedback and contributions. We also extend our gratitude to Dr. Jonghee Kim and Jaeyeong Jang for their guidance and insights, as well as to Co-walk Health Coop for their support in facilitating the study. This study was supported by the Korea SHE Foundation's 2025 Seed Grant for Researchers' Workshop and the National Research Foundation of Korea (NRF) (RS-2023-00262527).
\end{acks}

\bibliographystyle{ACM-Reference-Format}
\bibliography{reference}



\end{document}